\documentclass{emulateapj}
\usepackage{natbib}
\input{epsf.sty}

\newcommand{\St}{\mathrm{St}}

\begin{document}
\title{THE PHYSICS OF PROTOPLANETESIMAL DUST AGGLOMERATES. X. HIGH-VELOCITY COLLISIONS BETWEEN SMALL AND LARGE DUST AGGLOMERATES AS GROWTH BARRIER}

\shorttitle{Collisions of Agglomerates}
\author{\scshape Rainer Schr\"apler\altaffilmark{1}, J\"urgen Blum\altaffilmark{1}, Sebastiaan Krijt\altaffilmark{2,3} and Jan-Hendrik Raabe\altaffilmark{1}}
\altaffiltext{1}{Institut f\"ur Geophysik und extraterrestrische Physik, Technische Universit\"at Braunschweig, Mendelssohnstr. 3, D-38106 Braunschweig, Germany}
\altaffiltext{2}{Department of the Geophysical Sciences, The University of Chicago, 5734 South Ellis Avenue, Chicago, IL 60637, USA}
\altaffiltext{3}{Hubble Fellow}
\email{r.schraepler@tu-bs.de}

\shortauthors{Schr\"apler, Blum, Krijt \& Raabe}
\slugcomment{in print at ApJ}

\begin{abstract}

In a protoplanetary disk, dust aggregates in the $\mu$m to mm size range possess mean collision velocities of 10 to 60 m~s$^{-1}$ with respect to dm- to m-size bodies. We performed laboratory collision experiments to explore this parameter regime and found a size- and velocity-dependent threshold between erosion and growth.  By using a local Monte Carlo coagulation calculation and complementary a simple semi-analytical timescale approach, we show that erosion considerably limits particle growth in protoplanetary disks and leads to a steady-state dust-size distribution from $\mu$m to dm sized particles.


\end{abstract}
\keywords{methods: laboratory –- methods: numerical –- protoplanetary disks}

\maketitle

\section{INTRODUCTION} \label{kap:INTRO}

In the planetesimal-forming process, dust and ice particles grow from $\mu$m-sized grains to larger agglomerates. \citet{Weidi77} showed that the typical collision velocity of dust particles is directly connected to their sizes. Due to their different coupling times to the gas in the disk, dust agglomerates of different sizes gain relative velocities by drift motions and coupling to the gas turbulence. In the past decades, large efforts have been spent in the laboratory to investigate the outcomes of such collisions. It was shown that dust particles can easily grow from micrometers to about one centimeter in size \citep{Guettleretal2010,Zsometal2010} at 1 AU distance from the young star. At larger sizes, bouncing, fragmentation and erosion were found to dominate over growth in aggregate-aggregate collisions. However, a parameter range not yet investigated by experiments are the collisions between $\le 50 \mu$m-sized dust aggregates with dm- to m-sized bodies. This combination is relevant for the protoplanetary dust evolution, because recent investigations by \citet{Kothe2016} show that abrasion in bouncing collisions between cm-sized aggregates occurs, which produces small dust agglomerates. Therefore, we performed impact experiments with dust-aggregate projectiles of up to 50 $\mu$m in radius, impacting cm-sized dust targets at 20 to 60 m~s$^{-1}$.

The application of the results found in this study to the dust evolution in protoplanetary disks is based on our previous work \citet{SB2011}. There, we showed that impacts of micrometer-sized dust grains produce a cascade of particles that further erode the large dust aggregates until coagulation balances the erosive particle production. In the following, we will show that 2- to 50 $\mu$m-sized dust aggregates also contribute to the erosion of larger aggregates and that the erosive effect was even underestimated by \citet{SB2011}.

\section{EXPERIMENTAL APPROACH} \label{sect:EXACT}

\subsection {Particle Samples}\label{sect:EP}

For the dust-aggregate projectiles and targets, we used 1.5 $\mu$m-sized spherical particles consisting of white-colored (targets) and black-colored (projectiles) amorphous SiO$_2$. The material parameters of the white monodisperse particles have been described by \citet{BlumSchr2006}.  Unfortunately, no material properties for black-coated particles are available. However, as we will show below, the projectiles always completely stayed intact and stuck to the target so that net erosion was always caused by the white particles. Thus, modified cohesion between the black particles due to their coating should be rather unimportant for the erosional outcome.

\subsection {Experimental Setup}\label{sect:ESPL}

\begin{figure}
  \center
      \includegraphics[width=0.5\textwidth]{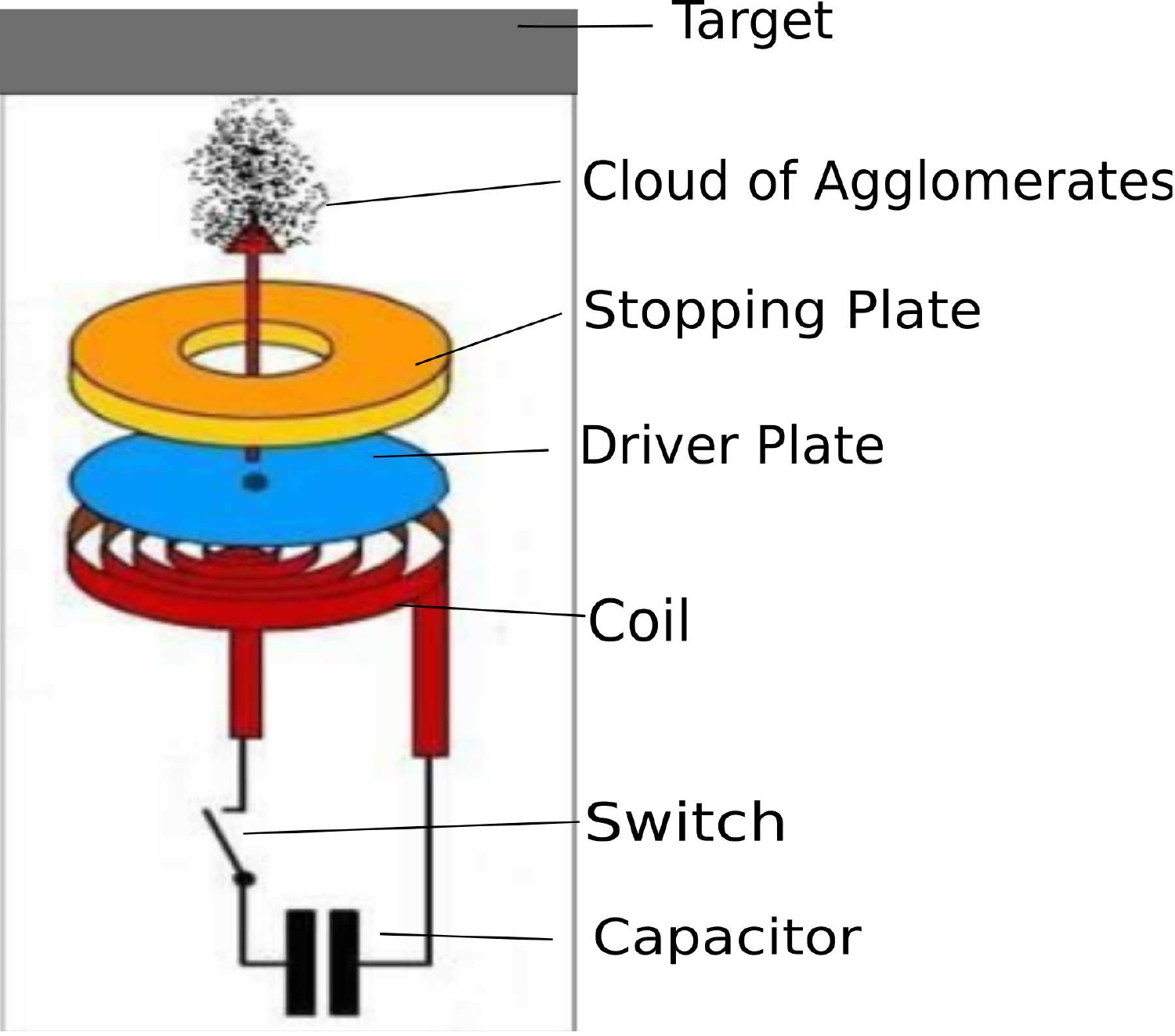}
    \caption{\label{fig:Coilgun} Schematic drawing of the experimental setup used in this study. It consists of a 5 $\mu$F capacitor that was charged for our measurements  with up to 9 kV. The capacitor is connected to a primary coil via a mercury switch. The emerging magnetic field induces an opposite field in the aluminum driver plate, which is being  accelerated upwards and stopped at the stopping plate. The projectile dust agglomerates, which are seeded on the aluminum plate (black dot in the center of the plate), detach from the driver plate when it impacts the stopping plate and continue flying through the hole in the stopping plate with up to 60 m~s$^{-1}$. A few mm above the stopping plate, the cm-sized dust-agglomerate target is mounted facing downward.
   }
 \end{figure}

\begin{figure}
  \center
      \includegraphics[width=0.5\textwidth]{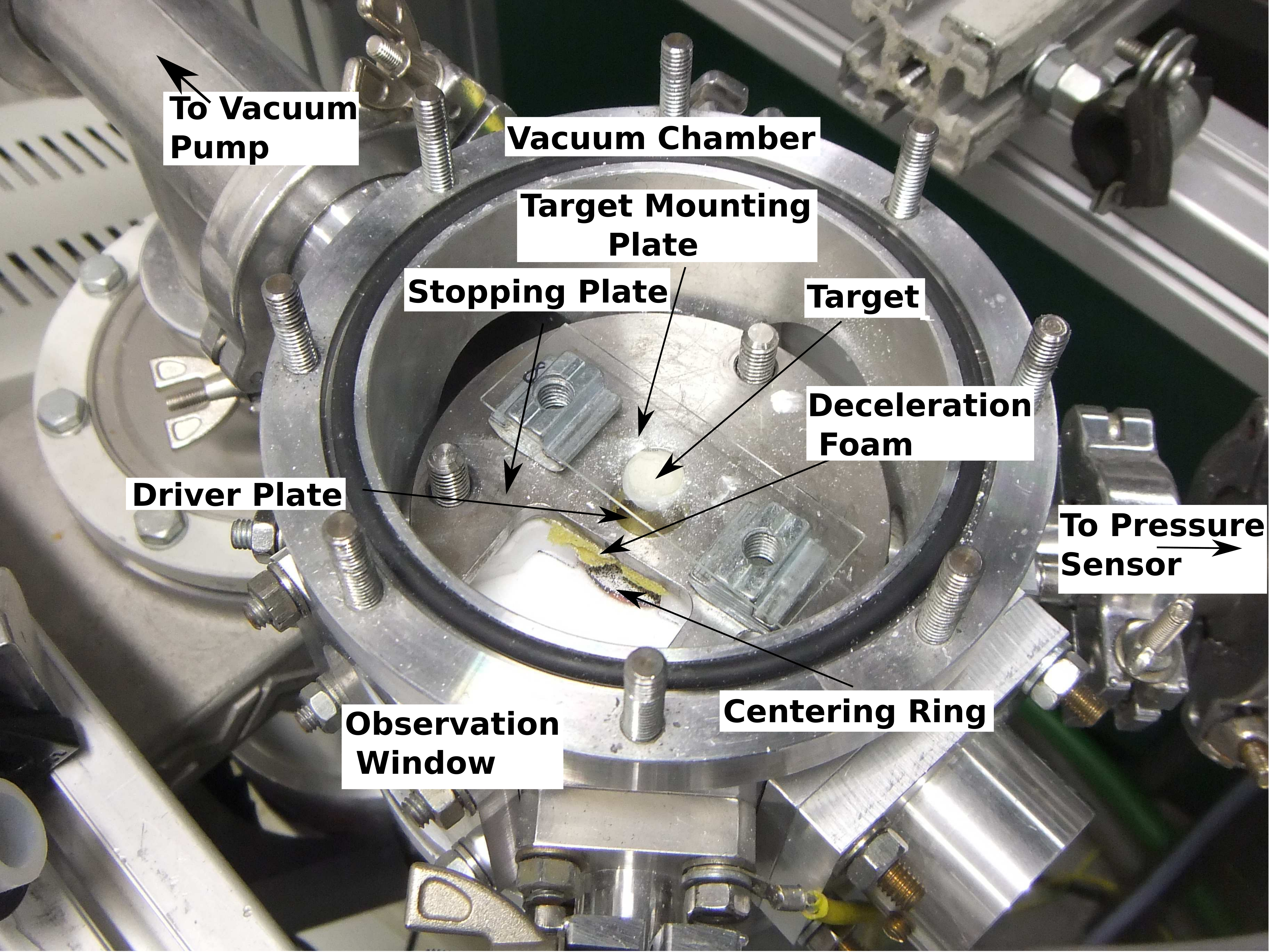}
    \caption{\label{fig:CoilgunEx}
  Photograph of the experimental setup. Close to the bottom of the vacuum chamber, the centering ring for the driver plate is visible. Below that, the driver coil is mounted (not visible). Above the driver plate, the stopping plate is mounted (note the hole in its center). On the bottom side of the stopping plate, a foam is glued to reduce the damage on the driver plate. The target is glued to a glass plate and mounted upside down over the hole in the stopping plate. On the lower left side, the vacuum chamber possesses an observation window.}
\end{figure}

The experimental setup used to produce impacts of small dust-agglomerate projectiles into cm-sized dust-agglomerate targets is shown in Figures \ref{fig:Coilgun} and \ref{fig:CoilgunEx}. An electromagnetic eddy-current accelerator is mounted inside a vacuum chamber, which is evacuated to a pressure of $< 1$~Pa. The coil gun consists of a 5 $\mu$F capacitor bank that can be charged up to 15 kV (outside the vacuum chamber). When the capacitor is discharged over the accelerator coil (primary spiral coil with 3 turns) via a mercury switch, the resulting rapid change of magnetic-field flux induces an eddy current and, thus, a magnetic field of opposite direction in the driver plate, which is initially resting on the primary coil. Due to the magnetic repulsion, the driver plate is accelerated upward and away from the primary coil. The driver plate possesses a diameter of 15 mm and a thickness of 0.3 mm and consists of the alloy AlCuMgPbF38. The dust-agglomerate projectiles are placed on the surface of the driver plate. At about one cm above the primary coil, the driver impacts a stopping plate. Hence, the dust agglomerates continue flying through a hole in the stopping plate and impact a cm-sized target agglomerate, which is glued on its upper end to a glass-plate holder. Thus, the target agglomerate is facing downwards a few millimeters above the hole in the stopping plate (see Figure \ref{fig:CoilgunEx}). A detailed description of the eddy-current accelerator including theoretical considerations is given in \citet{Lell1983}.

 Our projectile agglomerates possess compressive strengths of more than 1 MPa \citep{Machii2013}. Their tensile strengths are on the order of a few kPa \citep[see][]{BS2004}. For 60 ms$^{-1}$ impact velocity, the agglomerates are accelerated with $\approx 2\times 10^5$ ms$^{-2}$, which causes an internal pressure for our largest agglomerates of 4 kPa. This means none of our agglomerates is compacted during handling or acceleration. Agglomerates larger than 80 $\mu$m that are accelerated can break during acceleration to 60 ms$^{-1}$ or during handling. However, because the size of the projectile is measured after impact, potential break-up of the projectile can be taken into account. As the black aggregates did not break at impact, we can assume that their  strengths are at least not lower than the impact stress.

\subsection{Experimental Procedure}

\subsubsection{Projectile and Target Preparation}
The projectile agglomerates were produced by sifting the SiO$_2$ powder through sieves of 2$\mu$m to 35$\mu$m mesh widths. With this method, we produced dust agglomerates between $\sim 1$ and $\sim 50$ $\mu$m radius with a volume filling factor of $\Phi = 0.35$ \citep[see][]{Weidling2012}. A few hundred projectiles were then placed in the center of the driver plate using an orifice. Afterwards, the driver plate with the projectiles was loaded onto the primary coil (see Figure \ref{fig:CoilgunEx}).

The cm-sized targets were produced by omnidirectional compression of a pre-determined quantity of SiO$_2$ powder to a packing density of $\Phi = 0.4$. Targets were then glued to a glass plate and mounted upside down a few millimeters above the hole in the stopping plate (see Figure \ref{fig:CoilgunEx}).

\subsubsection{Velocity Calibration of the Accelerator}

\begin{figure}
  \center
      \includegraphics[width=0.5\textwidth]{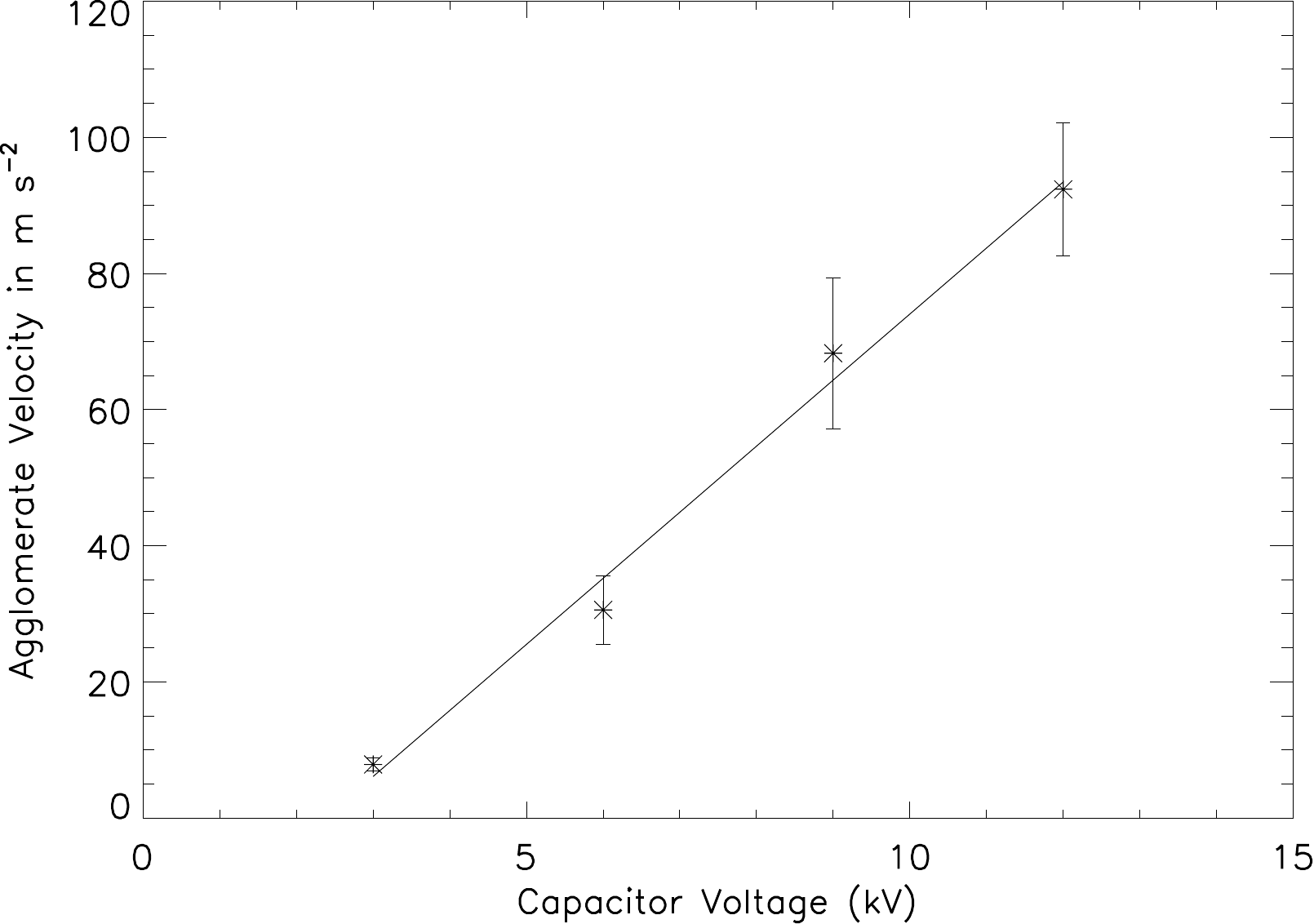}
    \caption{\label{fig:VolVel}  Velocities of dust agglomerates accelerated with the eddy-current device as a function of discharge voltage. The data points at 3 kV, 6 kV, 9 kV, and 12 kV are averages obtained from 19, 56, 18, and 18 individual velocity measurements, respectively. The error bars denote one standard deviation of the particle velocities. The solid line is a linear fit to the data and follows Equation \ref{eq:fit}.
 }
\end{figure}

To correlate the capacitor voltage with the velocities of the projectiles, we measured their speed by observing their flight paths with a high-speed camera through a window in the vacuum chamber (see Figure \ref{fig:CoilgunEx}). The result is shown in Figure \ref{fig:VolVel}. The symbols and errors correspond to the mean and standard deviation, derived by 19, 56, 18, and 18 velocity measurements for 3 kV, 6 kV, 9 kV, and 12 kV discharge voltage, respectively. A linear fit through the data yields a velocity-voltage correlation of
\begin{eqnarray}
\frac{v}{1~\mathrm{m~s^{-1}}} = 10 \times \frac{U}{1~\mathrm{kV}} - 25
\label{eq:fit}
\end{eqnarray}
which is shown by the solid line in Figure \ref{fig:VolVel}. Please note that the velocity of the particles is independent of their size.  A few hundred projectiles were positioned on the driver plate by sieving. It was not possible to further reduce this number. Therefore, we could not correlate each projectile sticking on the target with a projectile seen in the high-speed video images. Thus, the impact velocities were not obtained individually but by using Equation \ref{eq:fit}.

\subsubsection{Experimental Procedure}
For each individual experiment run, we followed the adopted protocol:
\begin{enumerate}
\item The target and driver plate are placed inside the vacuum chamber as described above.
\item The chamber is evacuated to better than 1 Pa pressure to avoid spark-overs and to minimize the gas-drag effects of the agglomerates.
\item The eddy-current gun is fired by discharging the capacitor over the primary coil so that the agglomerates impact the target at the pre-determied velocity (see Equation \ref{eq:fit}).
\item The chamber is vented.
\item The target is removed and analyzed under an optical microscope.
\end{enumerate}

\subsubsection{Target Analysis}
\begin{figure}
\center
      \includegraphics[width=0.5\textwidth]{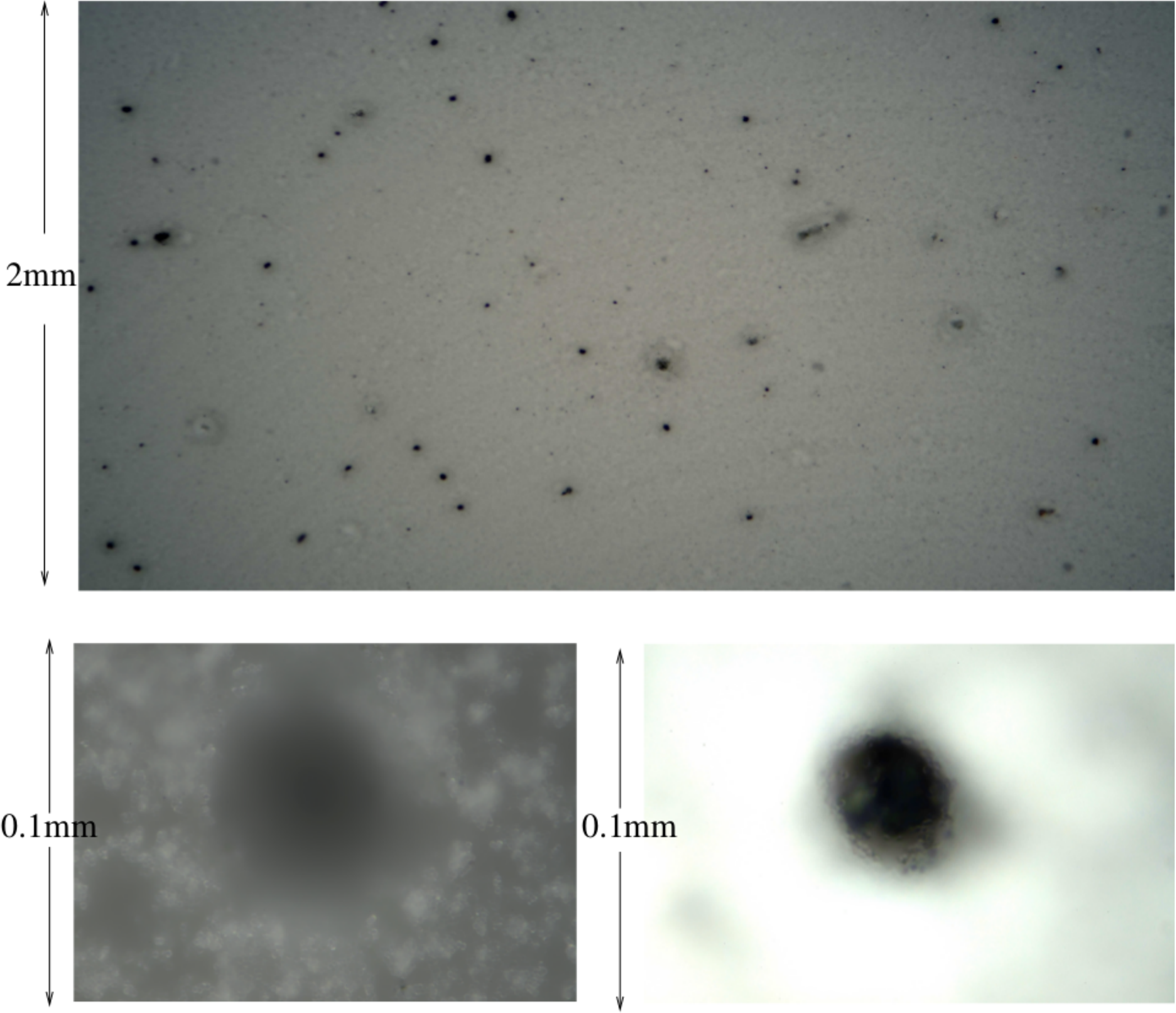}
    \caption{\label{fig:Tar}
    The upper picture shows the target consisting of white 1.5 $\mu$m-sized spherical SiO$_2$ particles (unresolved) to a packing density of $\Phi = 0.4$. Projectile-agglomerates with sizes of up to to $\sim 30~\mu$m consisting of black 1.5 $\mu$m-sized spherical SiO$_2$ particles are visible. The lower left picture is a high-resolution microscope image focused on the upper rim of a crater on the target surface, caused by an impact of a $\sim 30$ $\mu$m-sized projectile at 60~m~s$^{-1}$ impact velocity. Due to the low focal depth of the microscope, the black projectile is unfocused, but the diameter of the entrance hole of the crater can easily be measured. The lower right picture is focused on the bottom of the same crater with the black projectile inside. By adjusting the foci of these two images, the depth of the crater can be determined.}
\end{figure}

After the impacts, the target was removed from the vacuum chamber and placed under an optical microscope (Zeiss Axioskop). Here, the impactor sizes and the diameters of the impact craters on the target surface and at the depth at which the impactors stopped were measured (see Figure \ref{fig:Tar}). Because the depth of focus of our microscope was small against the intrusion depth of the projectiles, it was also possible to measure the crater depth (bottom pictures in Figure \ref{fig:Tar}). As our $\mu$m-sized grains were optically resolved by the microscope, we were able to estimate the filling factor on the crater walls and found that they were not compressed above $\Phi = 0.4$.

 Mounting the target upside down does not considerably influence our measurements. \citet{Meisner2013} performed experiments similar to ours and found in their Figure 10 that for impact velocities from 1 to 50 ms$^{-1}$ escaping fragments always have velocities larger than 20 cm s$^{-1}$, even at the lowest impact velocities within their error bars. Therefore, we are certain that particles that escape the target in our laboratory environment will also be lost under weightlessness conditions.

\subsection{Experimental Results}\label{sect:Expres}
We analyzed 153 impact experiments with projectile radii between $2 \rm ~\mu m$ and $30 \rm ~\mu m$ and impact velocities between $15 ~\rm m~s^{-1}$ and $60~ \rm m~s^{-1}$. A first general finding was that the impactors always stuck to the target, independent of their size and velocity. However, this does not mean that the target gained mass in the impact (see below).

\subsubsection{Penetration Depth}\label{sect:PD}
\begin{figure}
\center
      \includegraphics[width=0.5\textwidth]{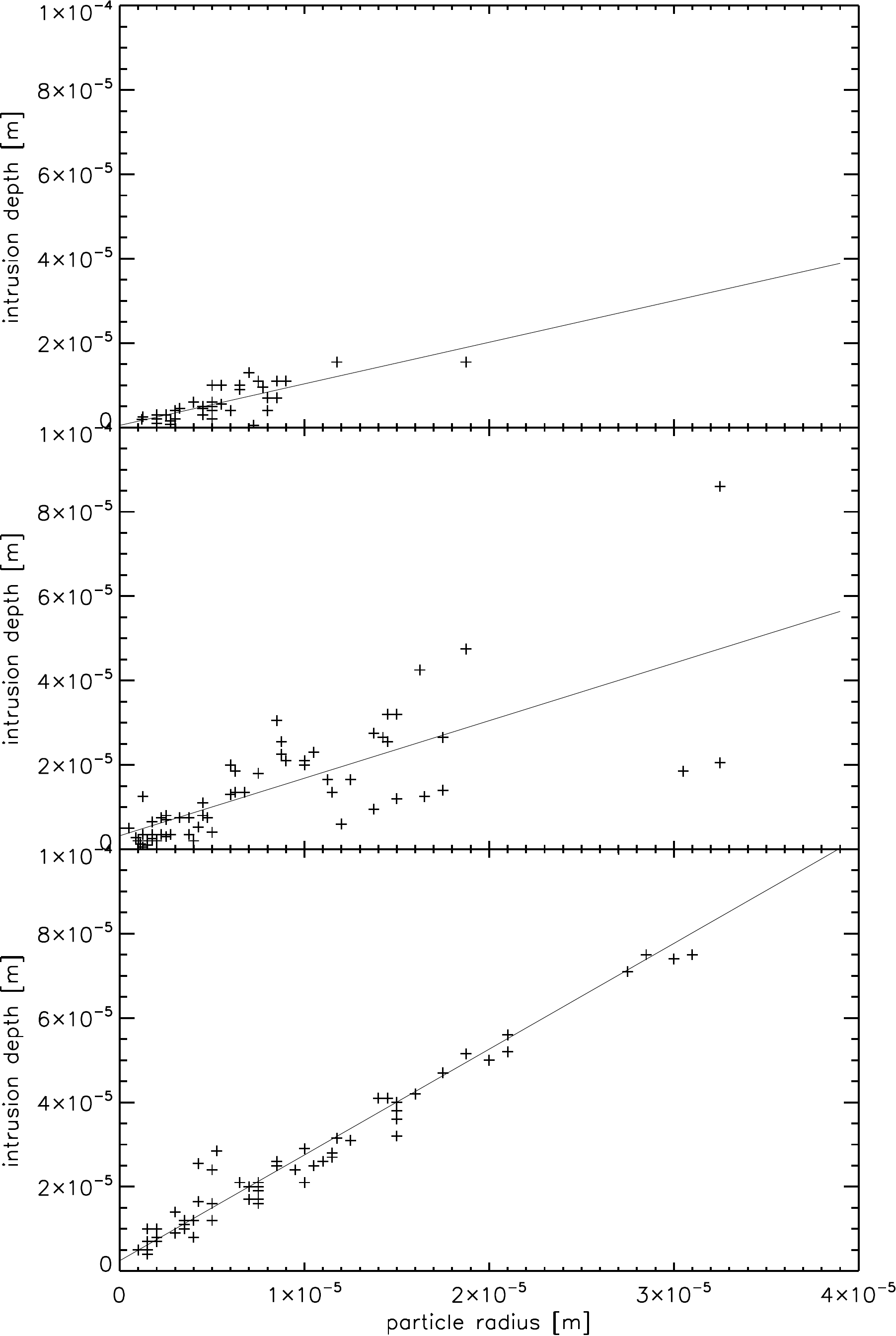}
    \caption{\label{fig:DoS} Penetration depths of projectile agglomerates with $\Phi$=0.35 impacting target agglomerates with $\Phi$=0.4 at impact velocities of $v\rm _{imp}=15~m~s^{-1}$ (top), $v\rm _{imp}=35~m~s^{-1}$ (middle) and $v\rm _{imp}=60~m~s^{-1}$ (bottom). The straight lines are linear fits to the data.}
\end{figure}

\begin{figure}
\center
      \includegraphics[width=0.5\textwidth]{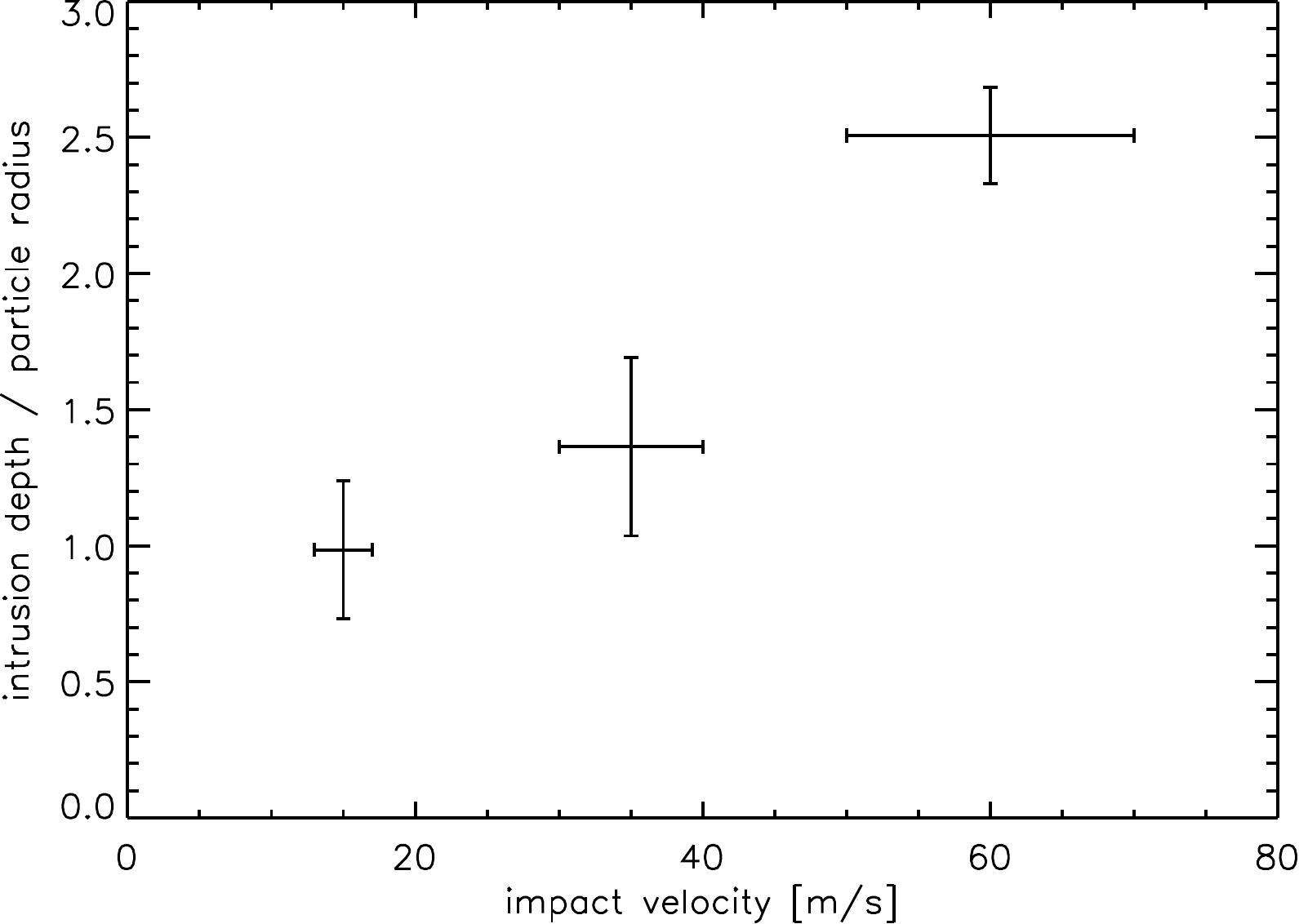}
    \caption{\label{fig:VeIn}Ratio between penetration depth and radius for dust agglomerates with $\Phi$=0.35 impacting target agglomerates with $\Phi$=0.4 as a function of their impact velocity. The data points were obtained through the fit lines shown in Figure \ref{fig:DoS}. The error bars denote two standard deviations.}
\end{figure}

Figure \ref{fig:DoS} shows the intrusion depth of the projectiles as a function of their radius for three impact velocities, $v\rm _{imp}=15~m~s^{-1}$, $v\rm _{imp}=35~m~s^{-1}$, and $v\rm _{imp}=60~m~s^{-1}$, respectively. One can clearly see a linear dependence between these two quantities, which is consistent with the findings of \citet{Guettleretal2009} for mm-sized solid projectiles impacting very fluffy dust targets at low velocities. We performed a linear fit to the data (solid lines in Figure \ref{fig:DoS}) and plotted their slopes and their 2$\sigma$ errors in Figure \ref{fig:VeIn}. As can be seen, the relative intrusion depth increases with increasing impact velocity. This also corresponds to the findings of \citet{Guettleretal2009} who measured a nearly linear dependency of the ratio between intrusion depth and projectile radius on the impact velocity. The finite y-axis intercepts ($<5\rm \mu m$) of the linear fits in Figure \ref{fig:DoS} presumably are due to a systematic error caused by the surface-roughness of the targets and hereafter ignored.

\subsubsection{Target Erosion}\label{sect:CE}
In a first attempt to investigate whether target erosion in impacts with dust-aggregate projectiles occurred at all, we covered the target in a closed box with a small hole of radius 0.5 mm through which the projectiles were injected at 60~m~s$^{-1}$. With this setup, we counted the impactors sticking on the surface of the target (94 particles between 25 and 35 $\mu$m and 26 particles between 35 and 50 $\mu$m in radius). This led to the estimation of the total impactor mass of 2.0 $\times 10^{-9}$ kg. To determine the ejected mass, we counted the particle debris found in the box (1771 particles between 0 and 5 $\mu$m radius, 884 particles between 5 and 15 $\mu$m radius and 324 particles between 15 and 25 $\mu$m radius) from which we derived a total ejected mass of 2.7 $\times 10^{-8}$ kg. Thus, the mass ratio between eroded mass (mass lost by the target) and impactor mass (mass gained by the target) is ${2.7 \times 10^{-8}}/{2.0 \times 10^{-9}}=14>1$ so that erosion was present. To estimate the maximum error caused by the binning of the sizes for impactors and ejected particles, we calculated the ratio of eroded mass and impactor mass by assuming that all impactors were at the maximum mass and all ejected particles were at the minimum mass of their respective bins. Even in that most pessimistic case, we get a mass ratio between eroded mass and impactor mass of 2.5 so that the presence of erosion is clearly proven.

To obtain quantitative results on the mass gain or loss of the target as a function of impactor mass and velocity, we determined the excavated mass from the crater in each individual experiment and compared it with the projectile mass that stuck to the target after the collision. We applied the following assumptions:
\begin{itemize}
\item The crater shape is a truncated cone (other shapes, e.g., a spherical calotte, do not change the result considerably) with the measured upper and lower diameters and height.
\item All matter below the cross section of the projectile is not ejected but compacted.
\end{itemize}


In case that the matter below the cross section of the projectile is also being ejected, the ejected mass would be a factor of $\sim 2$ larger and would better fit the result of our first attempt. However, due to the huge error bar of the experiment in the first attempt, we decided to remain with the more conservative ansatz and assume that the material under the projectile is not being emitted. Furthermore, a factor of $\sim 2$ in erosion strength would only marginally change the results of this paper.

\begin{figure}
\center
      \includegraphics[width=0.5\textwidth]{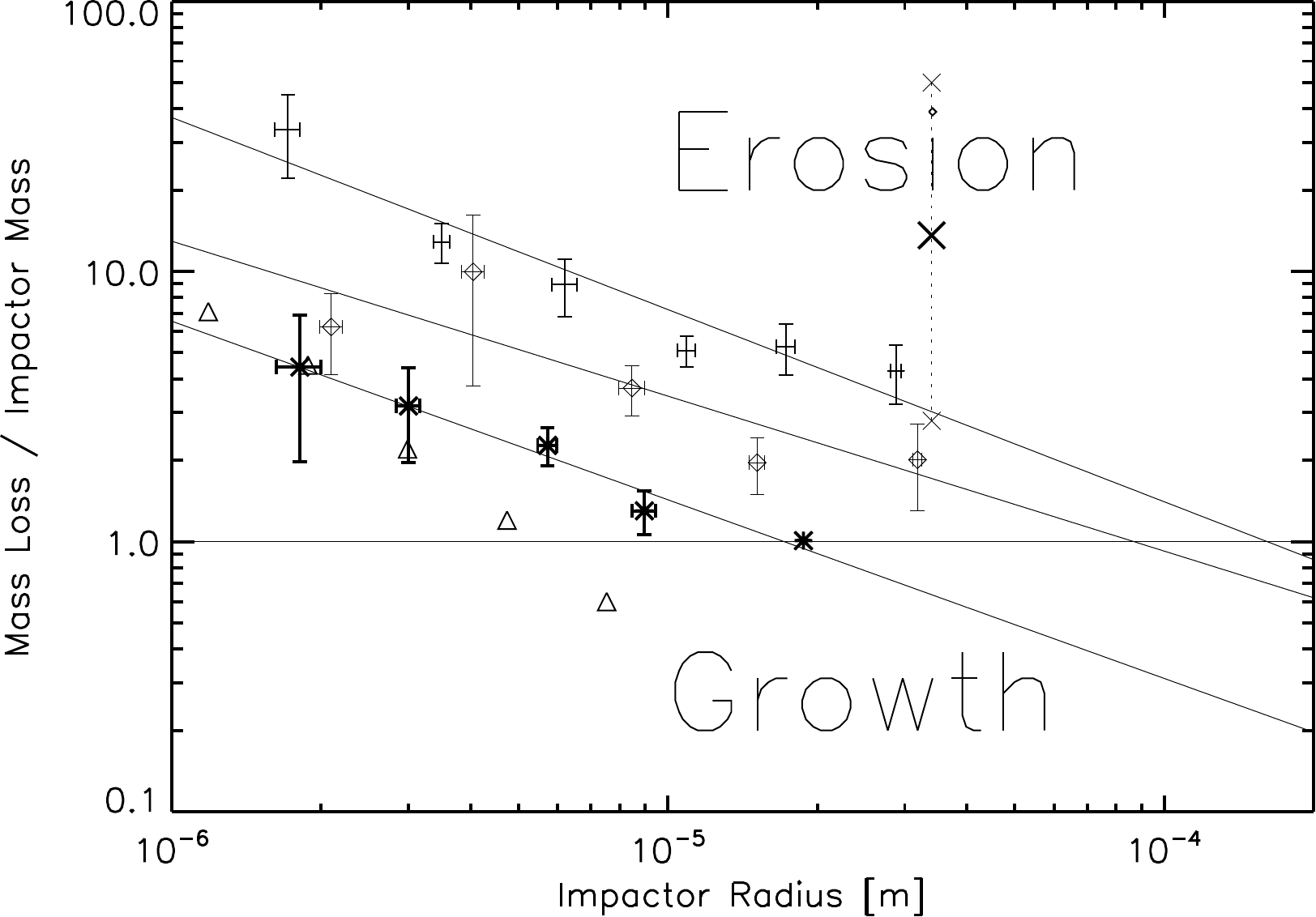}
    \caption{\label{fig:EOSII} Ratio between eroded mass and impactor mass as a function of impactor radius for impact velocities of 60 m~s$^{-1}$ (no symbols), 35 m~s$^{-1}$ (diamonds), and 15 m~s$^{-1}$ (asterisks), respectively. The triangles are taken from the numerical simulations of \citet{Seizingeretal2013}, who studied  impacts of projectiles into targets with both having a packing density of $\Phi=0.19$ at impact velocities of 15 m~s$^{-1}$. The data point denoted by an x with vertical dotted error bars stems from the calibration experiment at 60 m~s$^{-1}$ (see text).}
\end{figure}

Figure \ref{fig:EOSII} shows the ratio between eroded mass and impactor mass as a function of the projectile radius for the three impact velcities of 60 m~s$^{-1}$ (no symbols), 35 m~s$^{-1}$ (diamonds), and 15 m~s$^{-1}$ (asterisks), respectively. The triangles are taken from the numerical simulations of \citet{Seizingeretal2013}, valid for impacts of dust-aggregate projectiles and dust targets with $\Phi = 0.19$ at an impact velocity of  15 m~s$^{-1}$. The error bars denote one standard deviation of the measurements. The 15 m~s$^{-1}$ data were derived from 4, 8, 18, 7 and 1 measurements (from left to right), the 35 m~s$^{-1}$ data were derived from 13, 10, 15, 14 and 3 measurements, and the 60 m~s$^{-1}$ data were derived from 7, 8, 14, 13, 13 and 5 measurements, respectively. The data in Figure \ref{fig:EOSII} clearly demonstrate that the erosion efficiency depends on the impactor radius, following a power law with an exponent of $-0.5 \pm 0.1$, and increases with increasing impact velocity. \citet{Seizingeretal2013} found, that the erosion depends on the radius with an exponent of -1.3 (see triangles in Figure \ref{fig:EOSII}). The different exponents are probably caused by the different packing densities of their collision partners or their randomized impact directions. However, despite the differences between the simulations and the experiments, both nevertheless also show that there is a reduction of growth and a transition from erosion to growth for increasing impactor masses (horizontal straight line in Figure \ref{fig:EOSII}).

\begin{figure}
\center
      \includegraphics[width=0.5\textwidth]{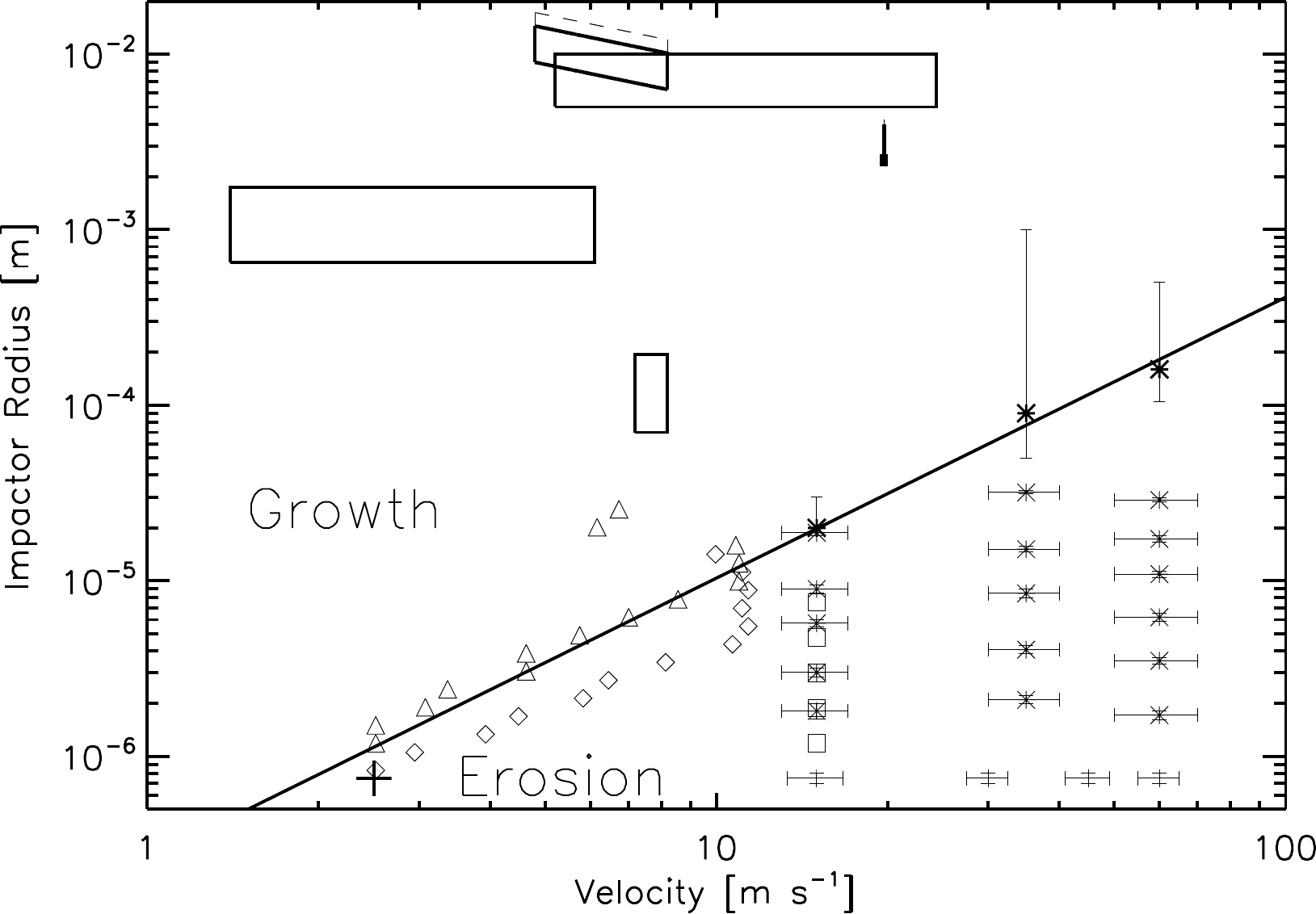}
    \caption{\label{fig:TV} Parameter range for growth and erosion covered by experiments and numerical simulations. Shown are the projectile radius on the y-axis and the collision velocity on the x-axis. The diagonal solid line separates the erosion regime (lower right) from the growth regime (upper left) and follows Eq. \ref{eq:f_eros}. The three asterisks with vertical error bars close to the solid line denote the threshold velocities and aggregate sizes derived in this study. The data from  Figure \ref{fig:EOSII} are shown by asterisks with horizontal error bars. The error bars with center cross show erosion from impact experiments of spherical monomers on targets with $\Phi$=0.15 \citep{SB2011}. The large plus sign shows the corresponding threshold velocity between erosion and growth for monomers taken from \citet{SB2011}. Triangles and diamonds denote the erosion-growth transition derived in the numerical experiments by \citet{Seizingeretal2013} for targets with $\Phi=0.19$ and $\Phi=0.55$, respectively. The squares without error bars show the corresponding numerical erosion runs from impacts of agglomerates of $\Phi$=0.19 on targets with $\Phi$=0.19 and $\Phi$=0.55 \citet{Seizingeretal2013}. In the growth regime, from top to bottom, the parameter space covered by the experiments of \citet{Deckers2014} ($\Phi=0.44$ for projectiles and target where the dashed area denotes onset of mass loss), \citep{Wurm2005} ($\Phi=0.34$), \citet{Teiser2011} ($\Phi=0.36$ for projectiles, $\Phi=0.33$ for target), \citet{Kotheetal2010} ($\Phi=0.15$)  \and \citet{Teiser2009} ($\Phi$ unknown) is shown.}
\end{figure}

The threshold velocity and impactor sizes above which growth occurs are found by extrapolation of the fit curves in Figure \ref{fig:EOSII} to the unity line. In Figure \ref{fig:TV}, we plotted these three values as asterisks. The error bars were derived by using extrapolations that just fit the measurements. The plus sign in Figure \ref{fig:TV} is the threshold velocity for the onset of erosion for $\mu$m-sized monomer particles found by \citet{SB2011}. We also added the numerical derivations of the erosion threshold from \citet{Seizingeretal2013} as diamonds (for $\Phi=0.55$ targets) and triangles (for $\Phi=0.19$ targets). The impactor agglomerates simulated by \citet{Seizingeretal2013} always have $\Phi=0.19$. As for larger impactors, their finite-size targets were disrupted before the erosion regime was reached  so that they were not able to perform calculations that are comparable to our measurements for projectiles consisting of more than 500 monomers (which corresponds to impactor radii larger than $6\times 10^{-6}$ m).

However, a clear power-law-type behavior of the relation between dust-agglomerate radius and threshold velocity for the onset of erosion is visible in Figure \ref{fig:TV} (solid line).

 Figure \ref{fig:EOSII} can be approximated by
\begin {equation}\label{eq:f_eros}
f=\left(\frac{r_{\rm{imp}}}{{2\times 10^{-5} ~ \mathrm{m}}}\right)^{-0.62} \frac{v}{15 ~ \mathrm{m ~ s^{-1}}}
\end{equation}
 By setting $f=1$, we obtain the separation curve between growth and erosion shown as a solid line in Figure \ref{fig:TV},
\begin {equation}\label{eq:r_imp}
\left( \frac{r_{\rm{thr}}}{1~{\mathrm{m}}} \right)= 2 \times 10^{-5} \left( \frac{v_{\rm{imp}}}{15~{\mathrm{m~s^{-1}}}} \right) ^{1.62}.
\end{equation}

 Erosion is present for impactor radii smaller than $r_{\rm{thr}}$ and growth is present for impactor radii larger than $r_{\rm{thr}}$.

We will come back to this point in Section \ref{sect:appli}. In addition, we mark the positions of previous collision experiments in Figure \ref{fig:TV}. All experiments to the right of and below the solid line in Figure \ref{fig:TV} led to erosion. Most of the measurements on the left of and above this line show mass gain of the target, independent of the packing densities. This confirms the presence of an erosion-growth boundary.

\subsubsection{Connection to other research}

\citet{Seizingeretal2013} found that passivation against erosion as described by \citet{SB2011} is an artefact of the persistent unidirectional impact of particles, whereas in a protoplanetary disk the rotational timescale of targets is much shorter than the impact frequency of projectiles so that unidirectional impacts cannot take place. 

\citet{Meisner2013} found growth within our erosive parameter range for particles about a factor of 5 smaller than those at our erosion-growth barrier. They also found stronger growth for smaller particles, which contradicts our ffinings. Their main conceptual difference to our work is that they did not perform measurements with fixed particle sizes. However, \citet{Meisner2013} measured the impact behavior of two samples of particles with wide size distributions. Both particle samples additionally possessed a wide velocity distribution whose correlation with the respective size distribution was not measured. In their paper, \citet{Meisner2013} state only the mean mass of the size distributions at a given mean velocity. We conclude that the origin of the discrepancy between our results and those of \citet{Meisner2013} occurs, because the erosional effect is different for different particle sizes and velocities. To theoretically find a proper mean particle size for their measurements with respect to the erosional effect, each particle size of their distribution function should be weighted with the respective erosional potential and impact velocity.


 \citet{Meisner2013} found growth within our erosive parameter range for particles about a factor of 5 smaller than those at our erosion-growth barrier. They also found stronger growth for smaller particles, which contradicts our findings. Their main conceptual difference to our work is, that they did not perform measurements with fixed particle sizes. However, \citet{Meisner2013} measured the impact behavior of two samples of particles with wide size distributions. Both particle samples additionally possessed a wide velocity distribution whose correlation with the respective size distribution was not measured. In their paper, \citet{Meisner2013} state only the mean mass of the size distributions at a given mean velocity. We think that the origin of the discrepancy between our results and those of \citet{Meisner2013} occurs, because the erosional effect is different for different particle sizes and velocities. To theoretically find a proper mean particle size for their measurements with respect to the erosional effect, each particle size of their distribution function should be weighted with the respective erosional potential and impact velocity.




\citet{Planes2017} did numerical investigations of agglomerate-projectile impacts into agglomerate targets, which is very similar to our experimental work but has partly substantially different results. The erosion yield obtained by \citet{Planes2017} is also directly proportional to the impact velocity of the projectiles and the ejected mass does not contain projectile material. Furthermore, their crater depth is proportional to their projectile size and the erosive effect is reduced for larger particles. However, the projectiles in the simulations by \citet{Planes2017} are completely destroyed upon impact, whereas ours stay intact. Their ejecta mass mostly consists of single monomer grains, whereas our mainly encompasses agglomerates with radii of about a quarter of the crater radius. The ejecta mass in \citet{Planes2017} is only 1\% to 10\% (depending on impactor size) of the mass  correspondent to the crater cavity volume. The remaining mass is stored within the target by compacting a volume 9 times the crater volume. This happens, because an extended volume (including the impactor) is fluidized upon impact, i.e. all bonds in between the monomers are broken.

The reason for these quantitative differences between the numerical simulations of \citet{Planes2017} and our experimental results are caused by substantially different values for the monomer-monomer break-up energy. \citet{Planes2017} used a value of $9\times 10^{-17}$ J for thebreak-up energy, whereas our particles possess a break-up energy of $9\times10^{-15}$ J. This means that \citet{Planes2017} simulated a much looser material, which is correspondingly easier to fluidize upon impact.


\section{Applications to Planetesimal Formation in Protoplanetary Disks: the Erosive Growth Barrier}\label{sect:appli}
In our previous work \citep{SB2011}, we showed that high velocity (15 m~s$^{-1}$ to 60 m~s$^{-1}$) impacts of micrometer-sized grains into large dust agglomerates lead to an erosion of these agglomerates. On the other hand, high velocity impacts of mm-sized dust aggregates cause a net growth of the target (see Figure \ref{fig:TV}). As shown above (see Figures \ref{fig:EOSII} and \ref{fig:TV}), we found an aggregate-size dependent velocity below which net growth and above which erosion dominate. Under protoplanetary disk conditions, erosion may potentially be important for projectile sizes between 1 and 100 $\mu$m.

 In the remainder of this Section, we study the effect of erosive collisions (as observed in Section \ref{sect:EXACT}) on aggregates growing in a protoplanetary nebula using two complementary methods: a local Monte Carlo coagulation calculation (Sect. \ref{sec:MC}) and a simple (semi)analytical model that compares timescales for growth, erosion, and radial migration of solids (Sect. \ref{sec:timescales}). The protoplanetary disk we consider has a gas surface density $\Sigma_\mathrm{g}(R) = 2000\mathrm{~g~cm^{-2}} \times (R/\mathrm{au})^{-1}$ and a temperature profile of $T(R)=280\mathrm{~K} \times (R/\mathrm{au})^{-1/2}$,  where $R$ represents heliocentric distance. A canonical dust-to-gas ratio of 1\% places the solid surface density at $\Sigma_\mathrm{d}=0.01 \Sigma_\mathrm{g}$.

\begin{figure*}[t]
\centering
\includegraphics[clip=,width=1.\linewidth]{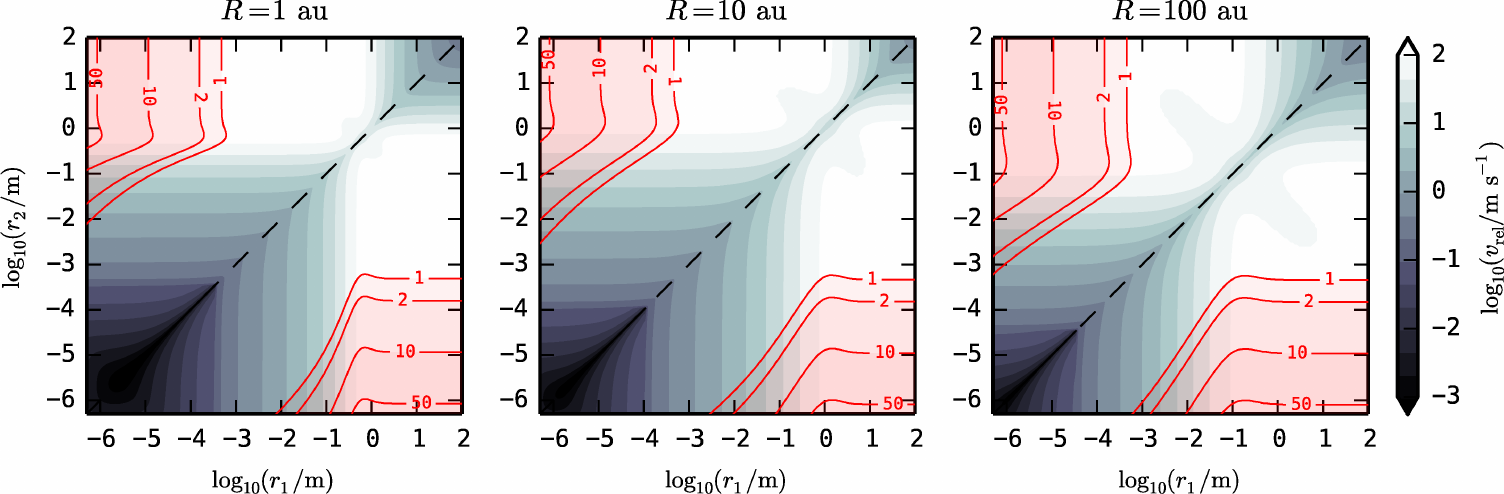}
\caption{The total relative velocity between grains of sizes $r_1$ and $r_2$ at three different heliocentric distances in the protoplanetary disk of Section \ref{sect:appli}. The dashed lines indicate same-size particles ($r_1=r_2$) and the red contours show where erosion takes place and what the erosion efficiency $f$ is according to Eq. \ref{eq:f_eros}.}
\label{fig:v_rel}
\end{figure*}

 Figure \ref{fig:v_rel} shows relative velocities, $v_\mathrm{rel}$, for pairs of particles with sizes $r_1$ and $r_2$ at three different heliocentric distances. When two particles collide, the impact velocity can be taken to equal $v_\mathrm{imp}=v_\mathrm{rel}$. The calculated relative velocities are dominated by interactions with the gaseous nebula and include contributions from Brownian motion, turbulence, radial drift, and azimuthal drift \citep[e.g.,][]{Weidi77,ormel2007b,okuzumi2012}. When computing $v_\mathrm{rel}(r_1,r_2)$, we have included both the Epstein and Stokes drag regimes, assumed a dimensionless $\alpha=10^{-3}$ when calculating the turbulent viscosity \citep[see][]{shakura1973,armitage2010}, and approximated the dimensionless pressure gradient as $\eta \approx 3 (c_s / v_\mathrm{K})^2$, with $c_s$ and $v_\mathrm{K}$ being the local sound speed and Keplerian orbital velocity \citep[]{armitage2010}. When calculating the dimensionless Stokes number ($\St$) of the aggregates, we assumed a dust particle density of $\rho_\bullet =1 \mathrm{~g~cm^{-3}}$, appropriate for a filling factor of $\Phi \sim 0.35$.

The red contours in Figure \ref{fig:v_rel} show the erosion efficiency $f$ (Eq. \ref{eq:f_eros}) and indicate a potential bottleneck for growth around a particle size of ${\sim}0.1{-}1\mathrm{~m}$ (corresponding to $\St\sim1$), at which point collisions with projectiles of sizes $r \lesssim 1 \mathrm{~mm}$ can result in erosion. In the next Sections, we will investigate whether the mass-loss of meter-size aggregates associated with this erosion can be severe enough to halt their further growth.

\subsection{Erosion in coagulation simulations}\label{sec:MC}
 First, we performed coagulation calculations using a representative Monte Carlo method, where all solids start out as microscopic grains with a diameter of $1\mathrm{~\mu m}$. The Monte Carlo method then calculates forward in time, using collision rates between representative particles to determine which grains collide \citep[see][for more details]{zsom2008,Zsometal2010,krijt2016b}. The simulations performed here are local in the sense that they are limited to a single heliocentric distance $R$, while the vertical dimension of the disk and the effect of settling of larger dust grains are taken into account when calculating the collision rates \citep[see also][]{okuzumi2012, krijt2015}.

To illustrate the potential power of erosion, we consider only the most optimistic scenario for growth in which the only outcomes of collisions are perfect sticking and erosion, for now ignoring catastrophic fragmentation and bouncing. When an erosive collision occurs, we assume that the formed fragments are distributed along an MRN power-law, with the largest possible fragment of a size comparable to that of the projectile. For a single collision, the total mass of fragments released equals $f$ times the mass of the impactor. The MRN distribution is the observational grain size distribution in the Milky Way \citep[see][]{Mathis1977}.

\begin{figure*}[t]
\centering
\includegraphics[clip=,width=1.\linewidth]{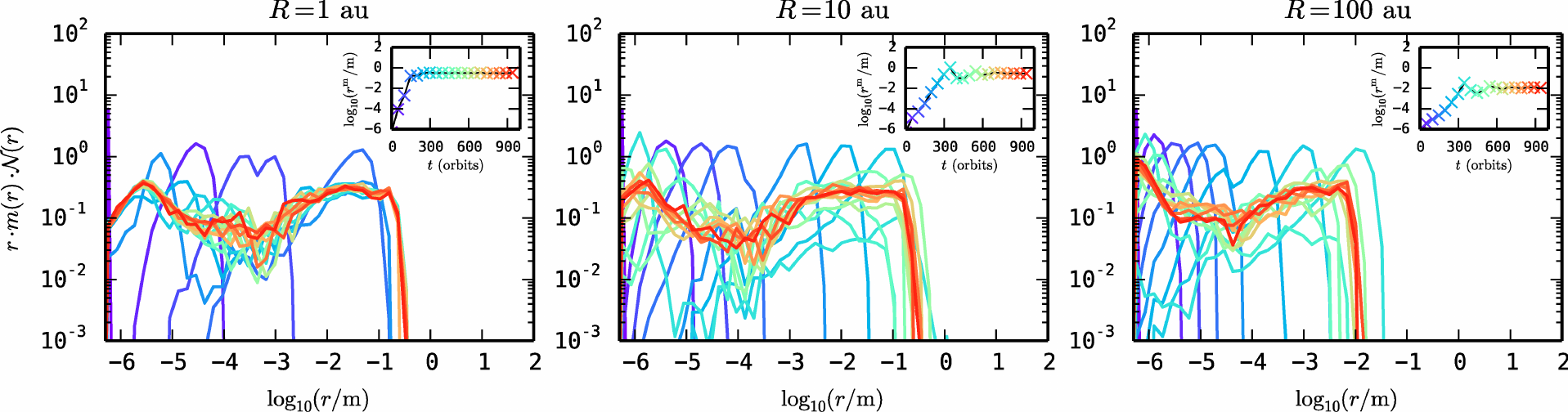}
\caption{Results of the coagulation models discussed in Section \ref{sec:MC} at 1, 10 and 100 au. The colored curves show normalized mass-weighted size distributions at different time steps (violet corresponds to $t=0$, red to $t=10^3$ orbits). The insets show the evolution of the maximum particle size $r^\mathrm{m}$ present in the simulation as a function of time.}
\label{fig:nama}
\end{figure*}

Figure \ref{fig:nama} shows the evolution of the dust size distribution over a time period of $10^3$ orbits for the same three locations that were shown in Figure \ref{fig:v_rel}, with each color representing the dust size distribution at a different point during the simulation. The quantity $\mathcal{N}(r)$ represents the column number density of particle of size $r$ and the $y-axis$ has been normalized so that the integral over the distribution equals 1. The insets in the top right show the maximum particle size at these same points in time. Qualitatively, the behavior is similar in all three cases: (i) Initially, a relatively narrow size distribution is growing to larger and larger sizes. During this period, the number of small grains is decreasing rapidly, as every collision results in sticking \citep{dullemonddominik2005}. (ii) As the largest particles begin to exceed sizes of ${\sim}\mathrm{cm}$, relative velocities become large enough for erosion to occur (see also Figure \ref{fig:v_rel}). As new fragments are created, the abundance of small grains starts to increase. (iii) For larger particles, the mass loss from erosive collisions is large enough to halt further growth. (iv) Eventually, a steady state situation is reached with part of the mass residing in sub-mm particles (red curves). The mass fraction in small grains exceeds the one found in growth-fragmentation steady-state distributions \citep[e.g.,][]{birnstiel2011} by several orders of magnitude.

\subsection{Comparing growth, erosion, and drift timescales}\label{sec:timescales}
As an alternative to the simulations of \ref{sec:MC}, we can also look at the importance of impact erosion by comparing the timescales for the ensuing mass loss to, for example, timescales on which radial drift or further growth take place. This approach is perhaps less accurate, as we do not solve for the abundance of small grains self-consistently, but is illustrative nonetheless and in good agreement with the results from the previous Section.

The assumptions we will make is that 50\% of the solid surface density $\Sigma_\mathrm{d}$ resides in large aggregates with size $r_L$, whose main mode of growth is collisions with roughly similar-size particles. The other 50\% of solids are `small' grains, with an MRN-like distribution between the monomer size $r_0=0.5\mathrm{~\mu m}$ and some maximum size $r_\mathrm{max} (\ll r_L)$. Using these assumptions, we can calculate characteristic timescales for erosion, growth, and radial transport for the large $r_L$ aggregates.

The radial drift timescale for the largest grains equals
\begin{equation}\label{eq:t_R}
t_R \equiv \frac{R}{|v_R|},
\end{equation}
where $v_R = - 2 \eta v_\mathrm{K} \St / (1+\St^2)$, which peaks around $\St \sim 1$ \citep{Weidi77}. Their growth timescale can be obtained as
\begin{equation}\label{eq:t_G}
t_G \equiv \frac{1}{\epsilon  \sigma n_L v_\mathrm{rel}},
\end{equation}
where $\sigma= \pi (2 r_L)^2$ is the collisional cross-section, $v_\mathrm{rel}$ will be dominated by the contribution of turbulence, and $n_L = 0.5 \Sigma_\mathrm{d} / (\sqrt{2\pi} H_L)$ is the midplane number density of large grains, which is found by assuming the grains have gravitationally settled into a layer with thickness $H_L \approx H_\mathrm{g} \sqrt{ \alpha / (\alpha + \St) }$, with $H_\mathrm{g}$ the gas pressure scale-height and $\St$ the Stokes number of the aggregates. The dimensionless efficiency factor $\epsilon$ is set to $\epsilon=0.5$. Approximating growth timescales by considering only the largest grains has been shown to be accurate when collisions predominately result in sticking \citep{krijt2016,sato2016}.

Finally, we define the erosion timescale as
\begin{equation}\label{eq:t_E}
t_E \equiv - \frac{m_L}{(\partial m_L / \partial t)_E},
\end{equation}
where $m_L$ is the mass of a single large aggregate and the mass-loss rate from erosion can be found by integrating the MRN distribution up to $r_\mathrm{thr}$,
\begin{equation}\label{eq:dmdt_E_1}
\left(\frac{\partial m_L}{\partial t}\right)_E = - \pi r_L^2 v_\mathrm{rel} \int_{r_0}^{r_\mathrm{thr}} n(r) m(r) (f-1) dr,
\end{equation}
 where we have used that the collisional cross section is dominated by the large grains, that $r_0 \leq r_\mathrm{thr} \leq r_\mathrm{max}$, and that the relative velocity does not depend on the size of the projectile (see Figure \ref{fig:v_rel}). From the previous Section we know that $r_\mathrm{thr} = r^* (v_\mathrm{rel}/v^*)^{1.62}$ and $f=(r/r^*)^{-0.62} (v_\mathrm{rel}/ v^*)$, with $r^*=2\times10^{-5}\mathrm{~m}$ and $v^*=15\mathrm{~m/s}$. Equation \ref{eq:dmdt_E_1} can be solved analytically if we assume that the erosion efficiency is large so that $(f-1)\approx f$. After plugging in $n(r) = n_0 r^{-q}$ for the MRN distribution\footnote{We will use $q=3.5$. The normalization constant $n_0$ can be found by integrating the mass density in the distribution and setting it equal to $0.5 \Sigma_\mathrm{d} /( \sqrt{2\pi}H_\mathrm{g})$.}, the particle mass $m(r)=(4/3)\pi r^3\rho_\bullet$ and $f=(r/r^*)^{-0.62} (v_\mathrm{rel}/ v^*)$, we obtain

\begin{equation}
\left(\frac{\partial m_L}{\partial t}\right)_E = - \mathcal{C}  r_L^2 v_\mathrm{rel}^2 \left[ \frac{r_\mathrm{thr}^{3.38-q} - r_0^{3.38-q}}{(3.38-q)} \right],
\end{equation}
where
\begin{equation}
\mathcal{C}=(4/3) \pi^2 \rho_\bullet n_0 (r^*)^{0.62} (v^*)^{-1}.
\end{equation}

\begin{figure*}[t]
\centering
\includegraphics[clip=,width=1.\linewidth]{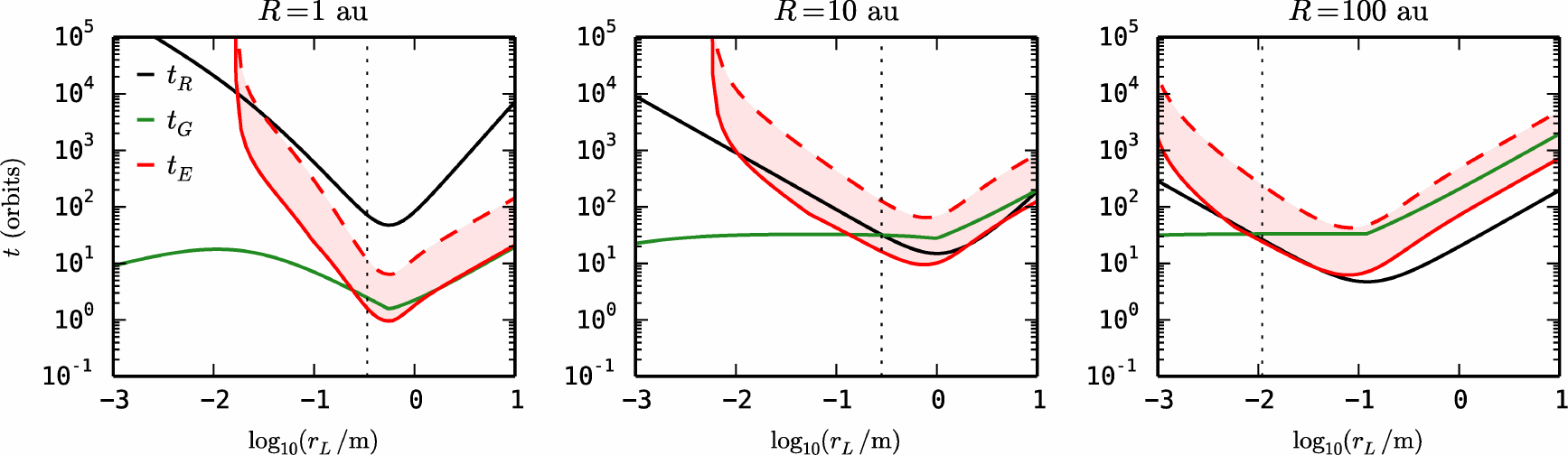}
\caption{Timescales for radial drift (Equation \ref{eq:t_R}, {\bf black line}), growth (Equation \ref{eq:t_G}, {\bf green line}), and erosion (Equation \ref{eq:t_E}, {\bf red lines, and red shaded area}) as a function of aggregate size $r_L$ (see Section \ref{sec:timescales}). The minimum in $t_R$ corresponds to a Stokes number of $\St\approx1$. Vertical dotted lines represent maximum particle sizes at the end of the simulations in Figure \ref{fig:nama}.}
\label{fig:timescales}
\end{figure*}

Figure \ref{fig:timescales} shows the drift, growth, and erosion timescales (normalized to the local orbital period) as a function of the large particle size $r_L$. Two curves are shown for $t_E$ to highlight the importance of the assumed small grain distribution: the solid line corresponds to $r_\mathrm{max}=10\mathrm{~\mu m}$ and the dashed line represents $r_\mathrm{max}=1\mathrm{~mm}$. It is evident that erosion is less efficient for  larger $r_\mathrm{max}$; the reason for this is two-fold: (1) the erosion efficiency (per unit projectile mass) decreases for increasing projectile size (see Figure \ref{fig:EOSII}); and (2) for a broader MRN distribution, the fraction of projectiles with $r>r_\mathrm{thr}$ (those that are excluded from the integral in Eq. \ref{eq:dmdt_E_1}) increases\footnote{In theory these larger projectiles should be included when calculating $t_G$, but for the parameters used here the effect is minor.}.

Focussing first on growth and drift, we see that for the smaller aggregates ($r_L \lesssim \mathrm{1~cm}$) the timescale for growth is shortest in all cases, indicating these particles will grow in-situ without significant radial drift taking place. As the size increases, radial drift becomes increasingly problematic for the 10 au and 100 au cases, showing the presence of the radial drift barrier. For the innermost case, however, growth is fast enough to break through the radial drift barrier - in part, because the aggregates are in the Stokes drag regime here \citep[see][]{okuzumi2012}.

As aggregates approach the size at which radial drift is fastest, i.e., $\St=1$, the relative velocity with the small grains also starts to increase (see Figure \ref{fig:v_rel}). As a result, the erosion timescale $t_E$ decreases significantly, and becomes comparable to the growth timescale. When the two are equal, further growth is not expected to take place. Indeed, the maximum particle sizes at the end of the simulations of Sect. \ref{sec:MC} (plotted as vertical dotted lines in Fig. \ref{fig:timescales}) line up well with the location where $t_G \approx t_E$.

\subsection{Implications for planetesimal formation \& avoiding the erosion barrier}
 Figures \ref{fig:nama} and \ref{fig:timescales} illustrate the findings of this Section: The erosion found in our experimental part (Equation \ref{eq:f_eros}) can single-handedly stop aggregates from growing through the drift barrier. When this occurs, a dust size distribution is created that contains abundant small fragments as well as a population of aggregates with a Stokes number just below 1. A similar conclusion was reached by \citet{krijt2015} for highly-porous aggregates growing beyond the water snow line, even though the physical size at which growth stopped was much larger because of the extremely low filling factors ($\Phi\sim10^{-5}$). While further growth through successive sticking collisions is hampered, the erosion barrier can in fact aid the formation of planetesimals via the streaming instability \citep[e.g.,][]{johansen2007N}, because it efficiently collects aggregates in the right Stokes number range of $10^{-2}< \St < 1$ \citep{carrera2015}.

A size distribution that is shaped by coagulation and erosion (i.e., the red curves in Figure \ref{fig:nama}) is not necessarily a universal outcome of dust evolution in protoplanetary disks. For example, if aggregates are easily disrupted and collision velocities are high enough, catastrophic fragmentation can stop growth at sizes well below a meter \citep[see][Figure 6]{birnstiel2011}. The same is true when bouncing is a dominant collisional outcome \citep{Zsometal2010}. Conversely, there are also scenarios where erosion around $\St \sim 1$ is less efficient and growth through the erosion barrier can still take place. For example, in a cooler disk $\eta$ will be lower, resulting in less efficient erosion because of the smaller $v_\mathrm{rel}$ (see Eq. \ref{eq:dmdt_E_1}). Alternatively, an extremely narrow size distribution (i.e., one in which there are virtually no small projectiles to kick-start the erosion process) could allow aggregates safe passage through the erosion barrier. However, this seems unlikely, because small microscopic grains are observed to exist in protoplanetary disks of all ages \citep{dullemonddominik2005}.

\section{Summary}\label{kap:COCON}
We experimentally investigated high-speed impacts of projectile agglomerates into cm-sized agglomerate targets and found that the erosion efficiency increases with increasing impact velocity and decreases with increasing projectile size. We found a transition from erosion to growth for a velocity-dependent projectile size.

 Finally, we used a local Monte Carlo coagulation model to investigate the impact of erosion on the dust distribution in a typical protoplanetary disk. We found that erosion halts growth at particle sizes of between one centimeter and a few decimeters, depending on heliocentric distance. A steady-state dust size distribution from micrometer to decimeter is maintained, which can explain observational results \citep{dullemonddominik2005}. We confirmed the erosional effect on dust growth with a complementary timescale-based analytical method.

\newpage



{\bf Acknowledgments:}
This research was supported by the Deutsches Zentrum f\"ur Luft- und Raumfahrt under grant nos. 50WM1236 and 50WM1536. SK acknowledges support from NASA through Hubble Fellowship grant HST-HF2-51394 awarded by the Space Telescope Science Institute, which is operated by the Association of Universities for Research in Astronomy, Inc., for NASA, under contract NAS5-26555.

\bibliographystyle{aa}
\bibliography{ms}

\newpage
\vskip5.0cm


\end{document}